\documentclass[twocolumn,showpacs,preprintnumbers,amsmath,amssymb]{revtex4}

\usepackage{graphicx}
\usepackage{dcolumn}
\usepackage{bm}
\usepackage{float}
\usepackage[normalem]{ulem}
\usepackage[usenames]{color}
\definecolor{Blue}{rgb}{0,0.0,1.0}
\definecolor{Red}{rgb}{1,0.0,0.0}
\definecolor{Grey}{rgb}{0.6,0.6,0.6}
\usepackage[utf8]{inputenc}
\usepackage[T1]{fontenc}

\begin{document}

\title{Quantifying reversibility in a phase-separating lattice gas: an analogy with self-assembly.}
\author{James Grant}
\email{R.J.Grant@bath.ac.uk}
\author{Robert L. Jack}
\affiliation{Department of Physics, University of Bath, Bath, BA2 7AY}


\newcommand{\comment}[1]{\textcolor{Red}{#1}}
\newcommand{\change}[1]{\textcolor{Blue}{#1}}
\renewcommand{\comment}[1]{#1}
\renewcommand{\change}[1]{#1}

\begin{abstract}
We present dynamic measurements of a lattice gas during phase separation, which we use as an analogy for
self-assembly of equilibrium ordered structures.  We use two approaches to quantify the degree of `reversibility' of this process:
firstly, we count events in which bonds are made and broken; secondly, we use correlation-response measurements and fluctuation-dissipation
ratios to probe reversibility during different time intervals.  We show how correlation and response functions can be related
directly to microscopic (ir)reversibility and we discuss time-dependence and observable-dependence of these measurements,
including the role of fast and slow degrees of freedom during assembly.
\end{abstract}

\newcommand{\ee}{\mathrm{e}}

\newcommand{\eb}{\epsilon_\mathrm{b}}
\newcommand{\ebT}{\eb/T}
\newcommand{\Teb}{T/\eb}

\newcommand{\from}{\leftarrow}

\maketitle

\section{Introduction}

Self-assembly~\cite{Whitesides2002-science,Glotzer2007,Halley2008} is the spontaneous formation of complex ordered equilibrium structures from simpler component particles.  
The range of possible structures includes novel crystals~\cite{Leunissen2005,Grzybowski2009,chung2010self,Romano2010}, 
viral capsids~\cite{Fox1998,Hagan2006}, and 
colloidal molecules~\cite{Wilber2007,Kraft2009,Sacanna2010}. 
Self-assembly is viewed as a promising alternative technology to current fabrication techniques, 
offering a bottom-up approach to design and manufacture~\cite{Ariga2008}.  
For example, experimental work has used DNA to generate a backbone for potential nanofabrication~\cite{Kershner2009,Rothemund2006}; potential applications of artificial viral capsids include inert vaccines and drug delivery~\cite{Brown2002}; and the potential range of self-assembled structures available through control of particle shape and interactions have been discussed 
extensively~\cite{Hagan2006,Glotzer2007,Wilber2007}.  
Self-assembly has even been considered for the potential regeneration of human organs and tissues~\cite{Stupp2010}.
However a key challenge for design and control of self-assembly processes is that even in systems where the
equilibrium states are known, the conditions which lead to effective assembly 
remain poorly understood and there remains no general theoretical approach to support experimental advances.

A major step towards such a theoretical approach has been the recognition of the importance of reversibility in the assembly 
process~\cite{Whitesides2002,Hagan2006,Jack2007,Rapaport2008,Whitelam2009}.  As a system evolves from an initially disordered state it makes bonds between the constituent particles.  If structures form which are not typical of equilibrium configurations then they must anneal before the system arrives at equilibrium.  When bonds between particles are too strong, thermal fluctuations are insufficient to break incorrect bonds before additional particles aggregate, and the system becomes trapped in long-lived kinetically frustrated states.  To avoid this problem and achieve effective self-assembly, bonds must be both made and broken as the system evolves in time.
In this sense, self-assembling systems are typically reversible on short time scales, even though they 
change macroscopically on long timescales.

Recent work \cite{Jack2007,Rapaport2008,Klotsa2011,Grant2011} 
has sought to quantify this qualitative idea by making dynamic measurements of reversibility.
We will compare two approaches in a simple lattice gas model where particles assemble into clusters and
the effectiveness of self-assembly is identified by the presence (or absence) of under-coordinated particles in the clusters.
Firstly, we count individual bonding and unbonding events and compare their relative frequencies by defining
dynamical flux and traffic observables~\cite{Grant2011}.
Then we use measurements of responses to an applied field to probe `reversibility', using fluctuation-dissipation 
theory~\cite{Cugliandolo1997,Cugliandolo1997a,Crisanti2003} combined with measurements of out-of-equilibrium correlation functions.  
We relate measurements of
fluctuation dissipation ratios (FDRs)~\cite{Crisanti2003} to the reversibility of self-assembly, extending the
analysis of~\cite{Jack2007,Klotsa2011}.  In particular, we derive
a formula for the response that elucidates its relation to microscopic reversibility and to measurements
of flux and traffic~\cite{Grant2011}.
We discuss how these results affect the idea~\cite{Jack2007,Klotsa2011} that measurements of reversibility on short
time scales might be used to predict long-time behaviour.

\section{The Model}
\label{sec:model}

\subsection{Definition}
\label{sec:defs}

We use an Ising lattice gas as a simple model system for self-assembly.  While this model appears much simpler than more detailed models of crystallisation~\cite{Leunissen2005,Romano2010} 
or viral capsid self-assembly~\cite{Hagan2006,Rapaport2008}, previous work has shown that
it mirrors many of the physical phenomena that occur in model self-assembling systems, 
particularly kinetic trapping~\cite{Whitelam2009,Hagan2011,Grant2011}. 

At low temperatures, the equilibrium
state of the lattice gas \comment{consists} of large dense clusters of particles, but efficient assembly of these clusters 
requires reversible bonding in order to avoid kinetic trapping (particularly aggregation into ramified fractal structures~\cite{Meakin1983}).

\begin{figure}
\includegraphics[width=8.2cm]{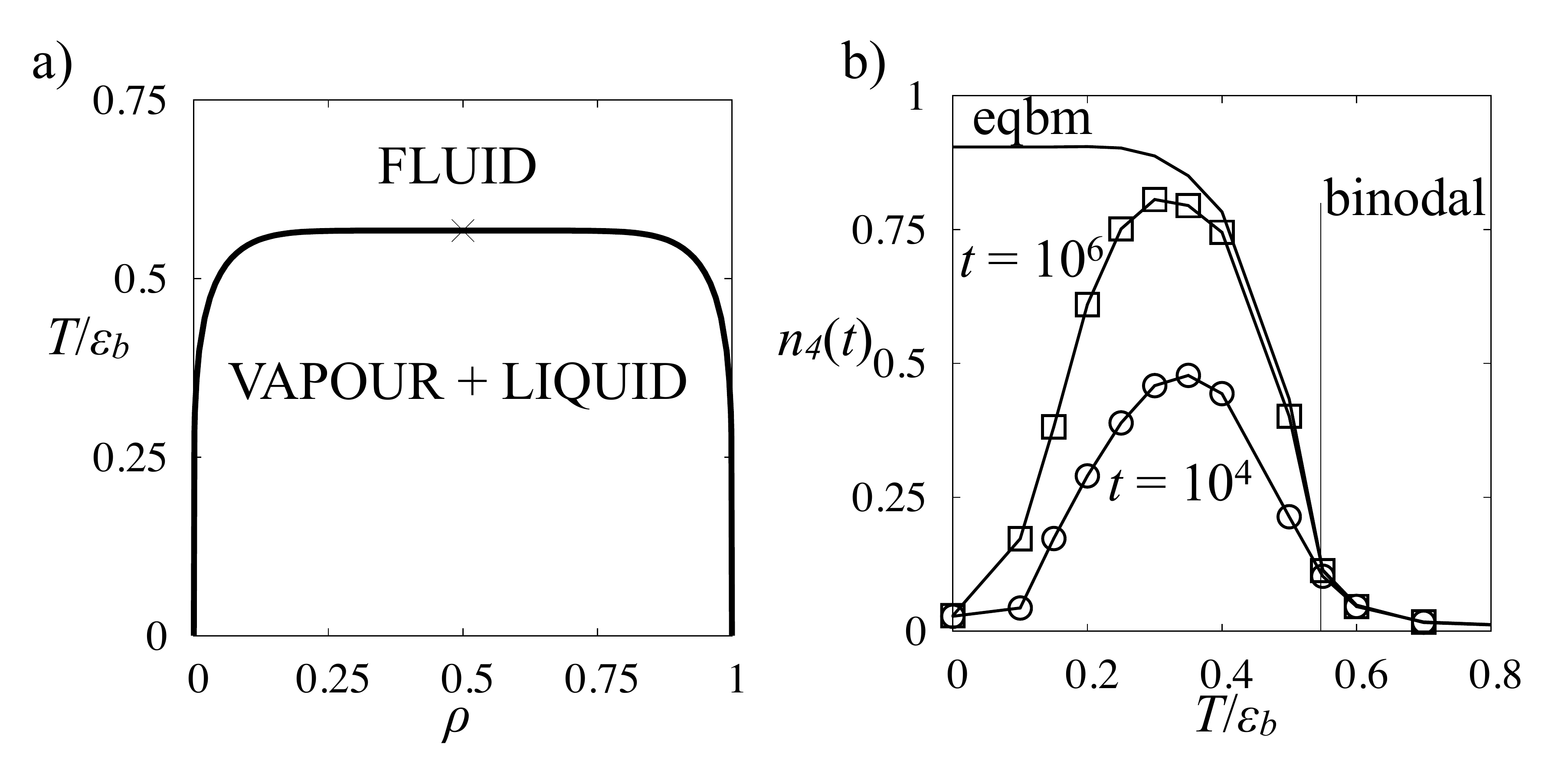}
\caption{a) Exact phase diagram for the lattice gas~\cite{Baxter2002}.  At high temperature the system equilibrates in a single fluid phase while below the binodal it separates into high and low density phases.  b) Plot of yield $n_4(t)$ against reduced temperature $\Teb$, for $t=10^4$ and $10^6$ MCS.  
For this
range of times, the yield is maximised at $\Teb\approx0.35$. The equilibrium yield is shown (labelled `eqbm'): as $t\rightarrow\infty$
the yield approaches this result. The binodal at $\Teb=0.547$ is shown as a vertical line.
}
\label{figyld}
\end{figure}

The model consists of 
a lattice, each site of which may contain at most one particle, and the energy of the system is
\begin{equation}
\displaystyle
E=-\frac{\eb}{2}\sum_p n_p
 \label{eqsysen}
\end{equation}
where the summation is over all particles, $\eb$ is the interaction strength, and
$n_p$ the number of occupied nearest neighbours of particle $p$.  We consider $N=1638$ particles on 
a square lattice of dimension $d=2$ and linear size $L=128$.  The relevant system parameters are the 
 density $\rho=N/L^d\approx 0.1$ and the bond strength $\eb/T$ (or equivalently, the reduced temperature $T/\eb$). 
The phase diagram is shown in Fig. \ref{figyld}, 
indicating the single fluid phase and the region of liquid-gas coexistence which occurs when 
$\sinh^{4}(\epsilon_{\rm b}/2T)>(1-(2\rho-1)^{8})^{-1}$\cite{Baxter2002}.
At the density $\rho=0.1$ at which we work, the binodal is located at $\Teb=0.547$, and
below this temperature the system separates into high and low density phases at equilibrium
(the critical point for the model is at $\rho_{c}=0.5$, $T_c/\eb=0.567$).  
\change{For the analogy with
self-assembly that we are considering, we are interested in behaviour at fairly low densities,
as in~\cite{Hagan2006,Wilber2007,Rapaport2008,Whitelam2009,Klotsa2011}.
All of the data that we show is taken at
$\rho=0.1$ but the same qualitative behaviour is found for lower densities too. (At higher densities,
we find that clusters of particles start to percolate through the system.  We do not consider
this regime since it would corresponds to gelation in the
self-assembling systems, and this is not relevant for the systems we have in mind.)}
 
We use a Monte Carlo (MC) scheme to simulate the diffusive motion of particles as they assemble. 
\change{Our central assumptions are (i) that clusters of $n$ particles diffuse with rates proportional to $1/n$,
and (ii) that bond-making is diffusion-limited (that is, bond-making rates depend weakly on the bond strength
$\ebT$ while bond-breaking rates have an Arrhenius dependence on $\ebT$.)}
Our simulations begin from a random arrangement of particles, and we simulate dynamics at fixed bond strength $\ebT$.
The dynamics are based on the `cleaving' algorithm~\cite{Whitelam2007} which makes cluster moves in accordance 
with detailed balance and ensures physically realistic diffusion.  
We propose clusters of particles to be moved by picking a seed particle at random: each neighbour of that 
particle is added to the cluster with probability 
$1-\exp\left(-\lambda\ebT\right)$
where $\lambda$ is a parameter that determines the relative likelihood of cluster rearrangement and cluster motion.
\change{We take $\lambda=0.9$ (see below).}
The cluster to be moved is built up by recursively adding neighbours of those particles in the cluster: each
possible bond is tested once.

In addition, a maximum cluster size $n_{\rm max}$ is chosen for each proposed move, and the move may be accepted only
if the size of the cluster to be moved is less than $n_\mathrm{max}$.  We choose
$n_\mathrm{max}$ to be a real number, greater than unity, distributed as $P(n_\mathrm{max}>n)=1/n^2$.
The result is that clusters of $n$ particles have a diffusion constant
$D\propto 1/n$, consistent with Brownian dynamics: see~\cite{Whitelam2007} for full details.

Having generated a cluster, 
one of the four lattice directions is chosen at random, and we attempt to move the cluster a single lattice spacing in that
direction, rejecting any moves which cause particles to overlap.   
Finally, the proposed cluster move is accepted with probability $P_{\rm a}$, which depends on its energy change $\Delta E$.
For $\Delta E\neq 0$, we take a Metropolis formula
$P_{\rm a}=\mathrm{min}\{1,\exp[(\lambda-1)\Delta E/T])\}$ while for $\Delta E=0$ we take 
\change{$P_{\rm a}=\alpha$ 
with $\alpha=0.9$ a constant (see below)}.
(Note that the definition of the energy in this model means that $\Delta E/\eb$ is an integer, which ensures that $P_\mathrm{a}$ is
monotonic in $\Delta E$ for all values of $\eb$ that we consider: cases where $\Delta E/\eb$ is
non-integer are discussed in Sec.~\ref{sec:resp-energy}.)
Simulation time, $t$, is measured in Monte Carlo sweeps (MCS), with $1\mathrm{MCS}$ corresponding to
$N$ attempted MC moves; the time associated with a single attempted MC move
is therefore $\delta t=1/N$. 
\change{We note that the constants $\alpha$ and $\lambda$ within this algorithm may be chosen freely between 0 and 1, while still
preserving detailed balance.  Our MC algorithm mimics diffusive cluster motion most closely when $\alpha$ and $\lambda$ are close to unity.
However, when evaluating response functions (see below),
the MC transition rates should be differentiable functions of any perturbing fields: this may not be the case 
if either $\alpha$ and $\lambda$ is equal to unity.
We therefore take $\alpha=0.9$ and $\lambda=0.9$.}

\subsection{Effectiveness of self-assembly}
 
\begin{figure}
\includegraphics[width=8.2cm]{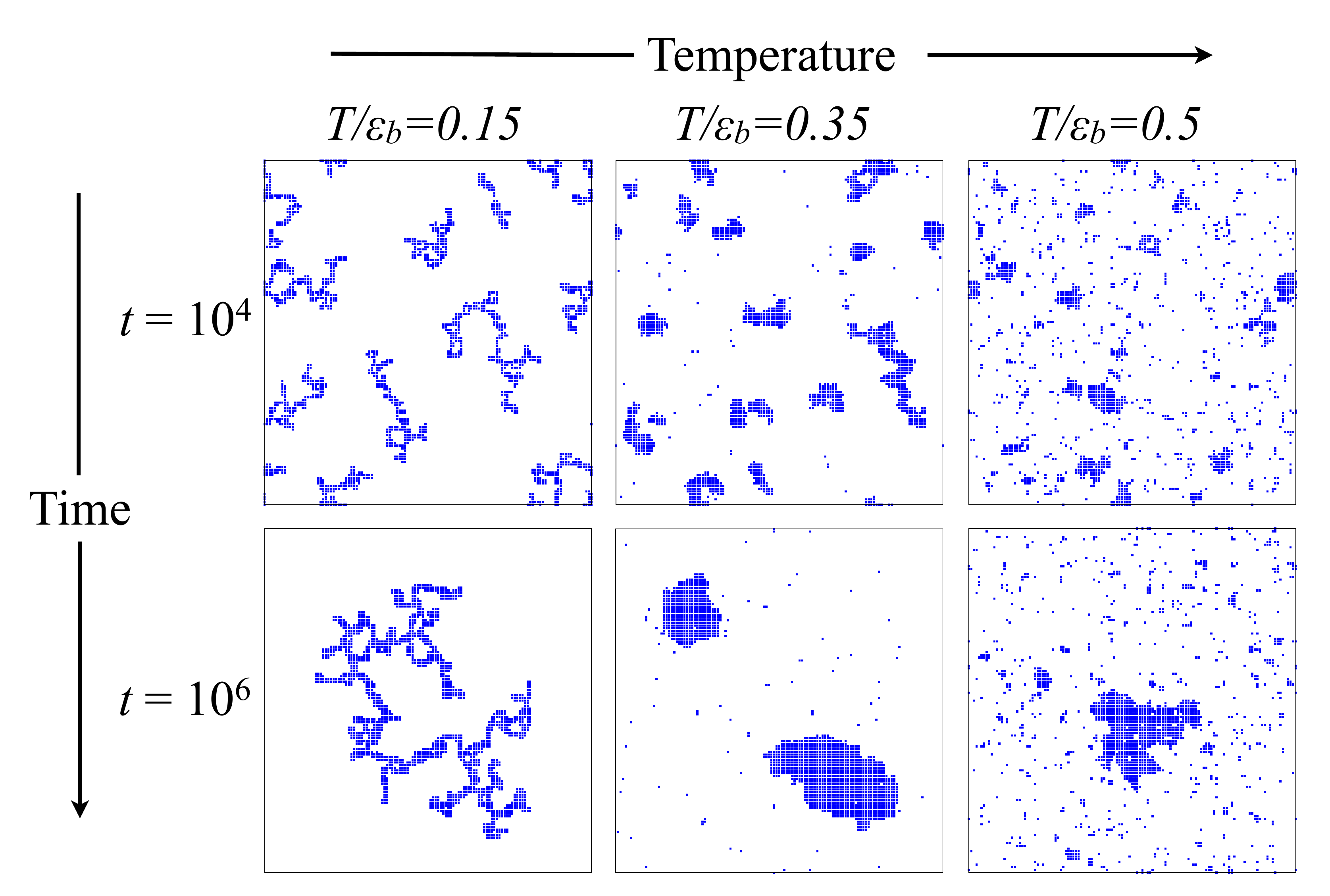}
\caption{
\change{
Snapshots of configurations at three representative temperatures. At $\ebT=0.15$ (low temperature), long-lived fractal clusters survive to form a stable gel-like structure, which we identify as kinetically frustrated.  At $\ebT=0.5$ (a higher temperature) and for long times, the system fully phase separates into a large cluster surrounded by a dilute gas.  However, the large cluster is not close-packed and a large proportion of particles remain in small clusters: this is therefore poor assembly.  At $\ebT=0.35$ (intermediate temperature) the majority of the particles are in large clusters having few defects, which we identify as good assembly.}%
}
\label{figconfig}
\end{figure}

\begin{figure*}
\includegraphics[width=16cm]{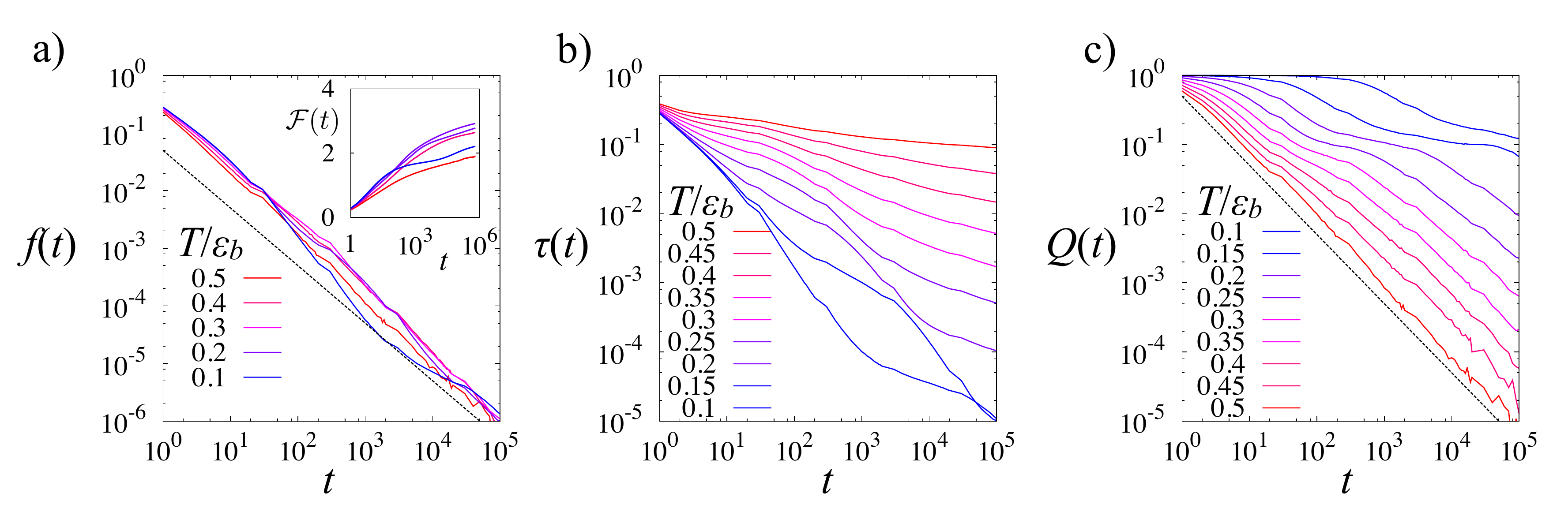}
\caption{Plots of the flux $f(t)$ and traffic $\tau(t)$, and their ratio $Q(t)$.  The flux varies little across the temperature range considered while the traffic depends much more strongly on the temperature.  In the kinetic trapping regime $\Teb\lesssim0.2$, the ratio $Q$ remains relatively large throughout the trajectories, while the good assembly and high-temperature regimes are associated with reversible evolution
and small values of $Q$.}
\label{figft}
\end{figure*}
 
In studying the lattice gas we have in mind a system self-assembling (or phase separating) 
into an ordered phase such as a crystal,
coexisting with a dilute fluid.
In the ordered phase, particles have a specific local co-ordination environment. In the lattice
gas model, we take this local environment to be that a particle has all neighbouring sites occupied. 
We define the `assembly yield', $n_4(t)$, as the proportion of particles with the maximal number of $4$ bonds to neighbouring particles.
In Fig. \ref{figyld}b we plot the average yield at two fixed times. 
We observe a maximum at a reduced temperature that we denote by $T^*/\eb$: the value of $T^*$ depends only weakly on time $t$.  
\change{The system shows three qualitative kinds of behaviour.  To illustrate this,
we show snapshots from our simulations in Fig.~\ref{figconfig}, at three representative bond strengths.
For strong bonds,
 there is a decrease in yield as the system becomes trapped in kinetically frustrated states. 
For weak bonds, the system is already close to equilibrium at $t=10^6$ but 
the final state has a relatively low value of $n_4(t)$.  
In the language of self-assembly, we interpret this as a 
poor quality product.  Based on Figs.~\ref{figyld} and~\ref{figconfig}, we loosely identify
the kinetically trapped regime as $\Teb<0.2$ and the regime of weak bonding and poor assembly as $\Teb\geq0.5$.  However,
we emphasise that these regimes are separated by smooth crossovers and not sharp transitions.  In the following,
we use $\Teb=0.15,0.35,0.5$ as representative state points for kinetic trapping, good assembly and poor assembly,
respectively.}

Thus, despite the simplicity of the lattice gas, it captures the kinetic trapping effects and the
non-monotonic yield observed in
more physically realistic model systems~\cite{Hagan2006,Wilber2007,Whitelam2009,Jack2007,Rapaport2008,Klotsa2011,Grant2011}.  
We emphasise that the kinetically trapped states we find are closely related to diffusion-limited aggregates~\cite{Meakin1983},
while the `assembled' states are compact clusters.  
\change{The changes in cluster morphology on varying bond strength
are discussed further in~\cite{Hagan2011} but in this article we concentrate on the dynamical
reversibility of assembly~\cite{Whitesides2002,
Jack2007,Rapaport2008,Klotsa2011} and not on the structure of the clusters.}
We include in Fig.~\ref{figyld} the equilibrium behaviour for systems of this size, which we obtain
by running dynamical simulations starting from a fully phase-separated state \change{in which a single
large cluster contains all particles}. As $t\rightarrow\infty$,
the yield must approach its equilibrium result: the laws of thermodynamics state that if we wait 
long enough then the highest quality assembly (or the lowest energy final state) will be at the lowest temperature.

\section{Measuring Reversibility}

\subsection{Flux-Traffic Ratio}
\label{sec:flux-traffic}

We now turn to measurements of reversibility and their relation to kinetic trapping effects and the non-monotonic
yield shown in Fig.~\ref{figyld}.
We follow~\cite{Grant2011} in considering the net rate of energy changing events at a microscopic level, the flux, in proportion to the total rate of energy changing events, the traffic (see also~\cite{Lecomte2007,Garrahan2009-kinks,Baiesi2009}).  
We count an event (or `kink') whenever a particle changes its number of neighbours: the number of times that particle $i$ increases its number of bonds between times $t$ and $t+\Delta t$ 
is $K^{+}_p(t,\Delta t)$, with $K^{-}_p(t,\Delta t)$ the number of times the particle decreases its number of bonds.
We then average and normalise by the time interval
\begin{equation}
\displaystyle
k^{\pm}(t,\Delta t)=\frac{1}{\Delta t} \langle K_p^{\pm}(t,\Delta t) \rangle.
 \label{eqkinks}
\end{equation}
\change{
Here and throughout, averages $\langle\dots\rangle$ run over the stochastic dynamics of the system
and over a distribution of initial configurations where particle positions are chosen independently at random.}
In the limit $\Delta t\rightarrow0$, then the $k^\pm(t,\Delta t)$ converge to the rates for bond-breaking 
($+$) and bond-making ($-$) events, which we denote by $k^\pm(t)$.
The flux, $f(t)=k^{+}(t)-k^{-}(t)$ is the net rate of bond-making and the traffic is the total
rate of all events $\tau(t)=k^{+}(t)+k^{-}(t)$.  
(In contrast to~\cite{Grant2011}, we focus here on rates for making and breaking bonds and not
on the total numbers of bonds made or broken.)

We also define a flux-traffic ratio
\begin{equation}
\displaystyle
Q(t)=\frac{f(t)}{\tau(t)}
 \label{eqftratio}
\end{equation}
which provides a dimensionless measure of the instantaneous reversibility of a system, 
consistent with the qualitative description of reversibility due to Whitesides~\cite{Whitesides2002}.  
The inverse of the flux-traffic ratio $1/Q(t)$ is the number of energy changing events per net bonding event. 
In a system at equilibrium $f(t)=0$ and so $Q(t)=0$ also. For a system quenched to $T=0$, our MC dynamics does not permit bonds to be broken so particles never increase their energy: hence $k^-(t)=0$ and $Q(t)=1$.  In systems that have been quenched, 
we expect to see a value between these two limits, $0<Q(t)<1$: 
the smaller the flux-traffic ratio the `more reversible' the system, while large flux-traffic ratios 
permit more rapid assembly.

In Fig. \ref{figft} we plot the flux $f(t)$, traffic $\tau(t)$, and their ratio $Q(t)$ at a range of bond strengths, $2<\ebT<10$.  
In all cases, the flux decreases towards zero as the system evolves towards equilibrium.  
On this logarithmic scale, the temperature-dependence of the flux appears quite weak, although the difference between
different bond strengths may be up to an order of magnitude.  
Given that the `integrated flux' ${\cal F}(t)=\int_0^t\mathrm{d}t' f(t')$~\cite{Grant2011}(Fig.\ref{figft}a inset)
 is always less than the maximal number of possible bonds
($4$ in this case), it is clear that $f(t)\lesssim t^{-1}$ at long times.


In contrast to the flux, the traffic shows a large variation with bond strength.  
Weaker bonds are more easily broken and result in more traffic.
Combining flux and traffic, the ratio $Q(t)$ is less than $0.1$ throughout the good assembly regime, and
decreases approximately as $t^{-1}$ while good assembly is taking place.  
The regime of kinetic trapping is characterised by larger values of $Q(t)$ and weaker time dependence.
Similar results were found in Ref.~\cite{Grant2011}, where a similar ratio denoted by $\tilde{M}$ was used
to compare total numbers of bond-breaking and bond-making events.  Compared to $\tilde M$, the ratio $Q(t)$ depends 
only on the state of the system at time $t$ and not on its history: this will be useful in making
contact with other measures of reversibility to be discussed below.  The 
ratio $1/Q(t)$ is in the range $10-1000$ during optimal assembly, indicating (as in~\cite{Grant2011}) that the system makes many steps
forwards and backwards before it achieves a single step of net progress towards the assembled state.
In this sense, good assembly requires significant reversibility, as argued above.

\subsection{`Flux relation' between correlation and response functions}
\label{sec:flux-relation}

In designing and controlling self-assembly processes, it would be useful if reversibility could be measured
and controlled, to avoid kinetic trapping effects.  While flux and traffic observables are not 
readily measured except in simple simulation models, we now show how correlation and response
functions can be used to reveal similar information (see also~\cite{Jack2007,Klotsa2011}).  
These functions can be measured 
without the requirement to identify specific bonding and unbonding events; in some cases
correlations and responses can also be calculated experimentally~\cite{Bonn07a, Ruocco10, Oukris10}.

At equilibrium, fluctuation dissipation theorems (FDTs)~\cite{Chandler1987} 
allow
the response of a system to an external perturbation to be calculated from correlation functions measured in the absence of the perturbation.
Away from equilibrium, such comparisons may be used for a classification of aging phenomena~\cite{Crisanti2003} and may also 
be useful for estimating the degree of (ir)reversibility in self-assembling 
systems~\cite{Jack2007,Klotsa2011}.

\begin{figure}[tt]
\includegraphics[width=8.2cm]{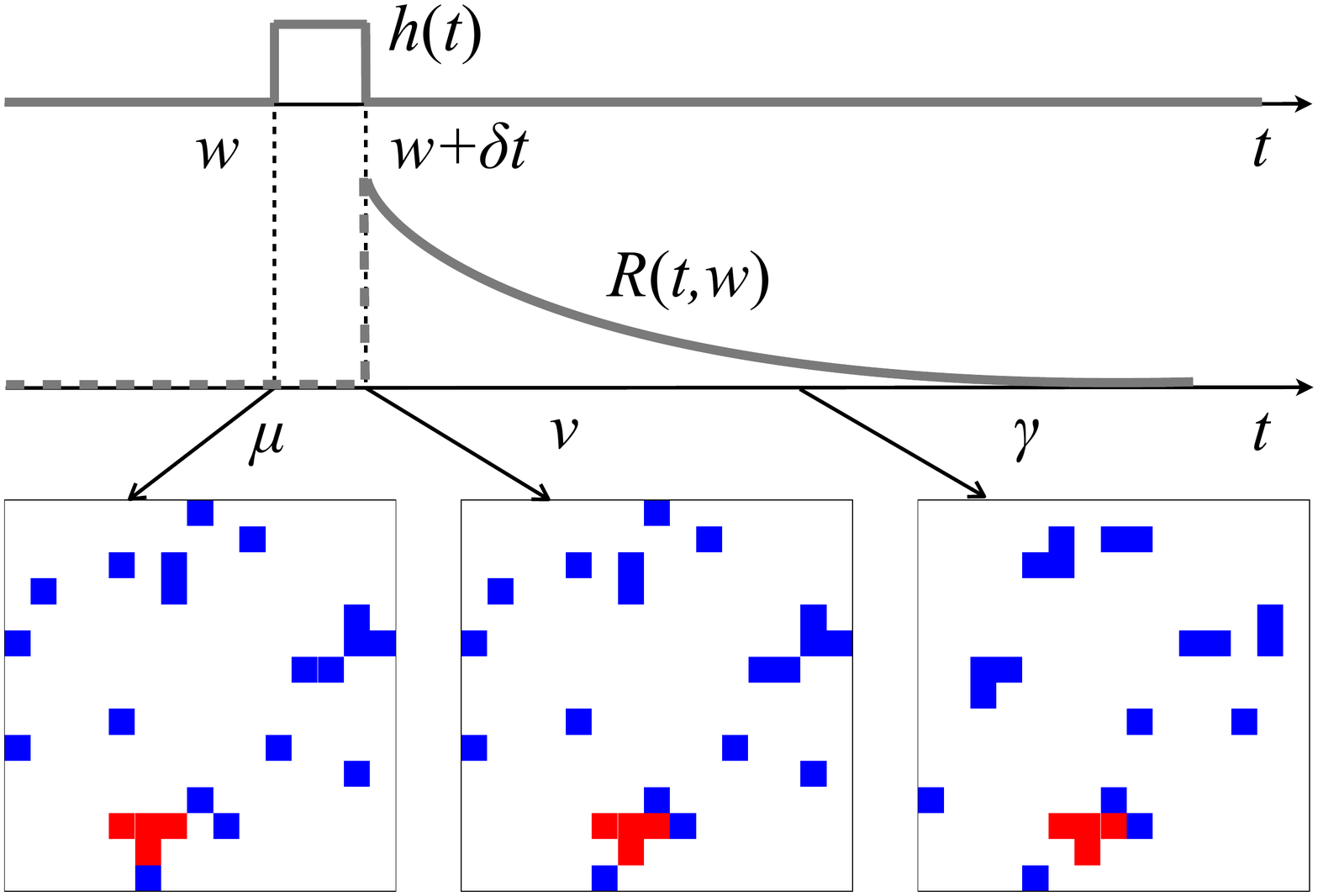}
\caption{
Procedure for measuring the impulse response. The system is initialised in a random configuration.  After a waiting time, $w$, a perturbation is
applied for a single MC move \comment{during which} the system attempts a move from configuration $\mu$ to $\nu$.  The field is then switched off and the system allowed to evolve until time $t$, at which point its configuration is denoted by $\gamma$.  The typical response of the system is indicated, together with snapshots of showing how particles might move through the system. (The cluster that moves in the step while the perturbation is applied is highlighted in red.)}
\label{figtimeline}
\end{figure}

\begin{figure*}
\includegraphics[width=16cm]{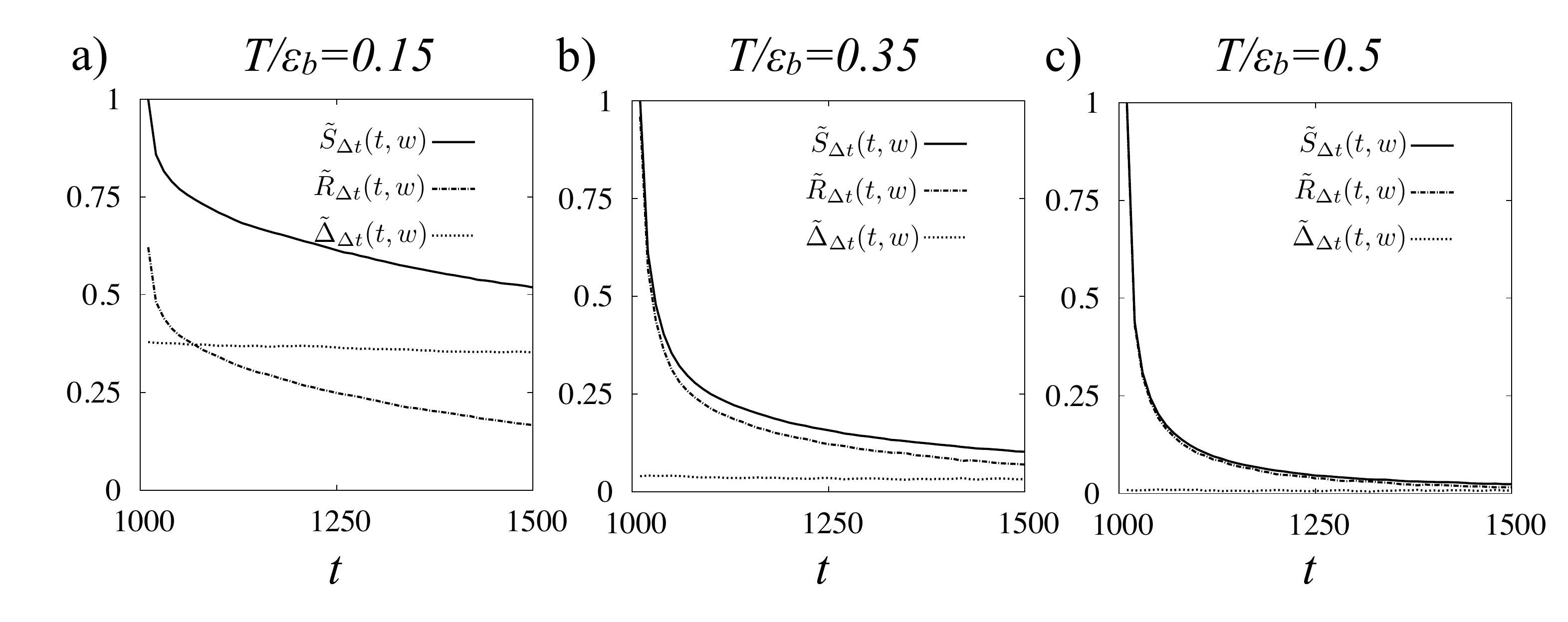}
\caption{
Data for $\tilde{S}_{\Delta t}(t,w)$, $\tilde{R}_{\Delta t}(t,w)$ and $\tilde{\Delta}_{\Delta t}(t,w)$ at three representative
temperatures.  The data are obtained at fixed $w=10^3$ MCS, while the time $t$ is varied.  The
deviation between correlation $S(t,w)$ and response $R(t,w)$ is small in both good assembly and poor assembly regimes,
but significant in the kinetically frustrated regime.  We emphasise that the correlation and response functions
are associated with perturbations to particle bond strengths: that is, $A=B=n_p$ in the definitions of $S$ and $R$.}
\label{figsr}
\end{figure*}

We first consider the general case for measurement of a response function in a system with MC dynamics.
Fig. \ref{figtimeline} illustrates our notation and the procedure used.  
The system is initialised at $t=0$ and 
evolves up to time $w$, when a perturbing field of strength $h$ is switched on.
A single MC move is attempted with the field in place, after which the field is switched off.
The system evolves with unperturbed dynamics up to a final time $t$ at which an observable $A$ is measured.
In general, the response function depends on the two times $t$ and $w$ and gives the change in the average
value of $A$, in response to the small field $h$.  As indicated in Fig. \ref{figtimeline}, we use Greek letters
to represent configurations of the system.  (Recall that $\delta t$ is the time associated with a single MC move.)

The MC scheme may be specified through the probabilities $P^0(\nu\from\mu)$ for transitions from configuration $\mu$ to $\nu$ in a single
attempted move.  However, it is convenient to work with transition rates
$W^0(\nu\from\mu)=(\delta t)^{-1} P^0(\nu\from\mu)$.   We also define $\rho_w(\mu)$ as the probability
that the system is in the specific configuration $\mu$ at time $w$, and the propagator $G^0_t(\nu\from\mu)$ as the probability 
that a system initially in state $\mu$ will evolve to state $\nu$ over a specified time period $t$.  
The superscripts $0$ indicate
that no perturbation is being applied to the system: we use a superscript $h$ if the field is applied.
\change{[It follows
that $\rho_w(\mu)=\sum_\kappa G^0_w(\mu\from\kappa)\rho_0(\kappa)$ where $\rho_0(\kappa)$ is the probability of initial condition
$\kappa$ in the dynamical simulations.]}

With these definitions, the average value of the observable $A$ at time $t$ is
\begin{equation}
\langle A(t) \rangle = \sum_{\gamma\mu\nu} A(\gamma) G^0_u(\gamma\from\nu) P^h(\nu\from\mu) \rho_w(\mu),
\end{equation}
where $A(\gamma)$ is the value of $A$ in configuration $\gamma$ and we introduce $u\equiv t-w-\delta t$ for compactness of notation.
(Since we use Greek letters to represent configurations,
there should be no confusion between $A(\gamma)$ and $\langle A(t)\rangle$, the former being a property of configuration
$\gamma$ and the latter a time-dependent average.)
The definition of the (impulse) response is 
\begin{equation}
R(t,w)\equiv\frac{T}{\delta t}\frac{\partial\langle A(t)\rangle}{\partial h},
\end{equation}
where the derivative is evaluated at $h=0$.  
Hence,
\begin{equation}
R(t,w)=T\sum_{\gamma\nu\mu}A(\gamma)G^0_u(\gamma\from\nu)\frac{\partial}{\partial h}W^{h}(\nu\leftarrow\mu)\rho_w(\mu).
 \label{eqr1}
\end{equation}

We assume that the system obeys detailed balance with respect to an energy function $E^h=E^0-hB$
where $E^0$ is the energy of the unperturbed system and
$B$ is the conjugate observable to the field $h$.
It is convenient to define a connected correlation function 
\begin{equation}
C(t,w)=\langle \delta  A(t) \delta B(w) \rangle
\end{equation} (here and throughout we use
the notation $\delta O = O-\langle O\rangle$ for all observables $O$). 
We also define 
\begin{equation}
S(t,w)\equiv\frac{\partial}{\partial w}C(t,w).
\end{equation}
[The MC dynamics evolves in discrete time steps: the interpretation of the time-derivative
is discussed in Appendix~\ref{app:response}.]  At equilibrium, one has the FDT
\begin{equation}
R^\mathrm{eq}(t,w) = S^\mathrm{eq}(t,w).
\label{equ:fdt}
\end{equation}
Out of equilibrium, we define 
\begin{equation}
\Delta(t,w) \equiv S(t,w) - R(t,w)
\label{equ:def-Delta}
\end{equation}
 as a deviation between the response
of the actual system and the response of an equilibrated system with the same correlation functions.


Various out-of-equilibrium expressions for response functions may be derived: see for 
example~\cite{Baiesi2009,Seifert2010,Corberi2010} where different representations are defined
and compared with each other.
In Appendix~\ref{app:dev}, we give a proof of (\ref{equ:fdt}) and we derive the general relation 
\begin{equation}
\Delta(t,w)=\sum_{\gamma\nu\mu} 
[A(\gamma) - \langle A(t)\rangle] G_u(\gamma\leftarrow\nu)
F(\nu,\mu) J^0_w(\nu,\mu)
 \label{eqdeviation}
\end{equation}
where 
\begin{equation}
F(\nu,\mu)=T\frac{\partial}{\partial h}\ln W^{h}(\mu\leftarrow\nu)
\label{equ:Fnm-general}
\end{equation} 
and
\begin{equation}
J^0_w(\mu,\nu) = W^{0}(\nu\leftarrow\mu)\rho_w(\mu)-
W^{0}(\mu\leftarrow\nu)\rho_w(\nu)
\label{equ:j0}
\end{equation}
is the probability current between configurations $\mu$ and $\nu$ at time $w$.
We will see below that the probability
current $J^0$ plays a central role in quantifying irreversibility and deviations from equilibrium.
%

\change{
The formula (\ref{eqdeviation}) is general for discrete time Markov processes obeying detailed balance (the generalisation
to Markov jump processes in continuous time is straightforward).  
We emphasise that (\ref{eqdeviation}) is a representation of (\ref{eqr1}), valid whenever detailed balance
holds.  As such, it is mathematically equivalent to several other representations that
have been derived elsewhere: for example, the `asymmetry' function discussed by Lippiello~\emph{et al.}~\cite{Lippiello2005,Corberi2010} 
plays a role similar to $\Delta(t,w)$ in the case of Langevin processes.  The purpose of (\ref{eqdeviation}) is to make contact
between reversibility and deviations from FDT in self-assembling systems, as we now discuss.  Further details of the relation between
(\ref{eqdeviation}) and previous analyses~\cite{Chatelain2004,Ricci-Tersenghi2003,Lippiello2005,Baiesi2009,Seifert2010,Corberi2010} are discussed in 
Sec.~\ref{sec:lippi-etc} below.
}

\subsection{Energy correlation and response functions in the lattice gas}
\label{sec:resp-energy}

We now turn to response functions in the lattice gas. 
We consider how a single particle in the lattice gas responds if the strength of its bonds are increased.
To this end,
we write the energy in the presence of perturbing fields $h_p$ as $E^h=\sum_p [\frac12\eb-h_p]n_p$ where the sum
runs over all particles, as in (\ref{eqsysen}).
We measure the response of $\langle n_p\rangle$ to the field $h_p$, by taking the observables $A$ and $B$ of the previous section
both equal to the number of bonds $n_p$ for a specific particle $p$.  Thus 
$R(t,w)=\frac{T}{\delta t}\frac{\partial \langle n_p\rangle}
{\partial h_p}$. Given
the energy function $E^h$, there is still considerable freedom to choose the MC rates $W^h$ while
preserving detailed balance.  

\change{In~\cite{Lippiello2005,Baiesi2009,Corberi2010}, it was mathematically convenient
to take $W^h(\nu\from\mu)=W^0(\nu\from\mu)\ee^{h[ B(\mu)-B(\nu)]/2}$.  Here our dynamics we are motivated by the 
central assumptions (i) and (ii) of Sec.~\ref{sec:defs}, which indicate that rates for bond-making and cluster diffusion
should depend only weakly on perturbations to the particle bond strengths.}
We therefore include all $h$-dependence in the probability $P_\mathrm{a}^h$ of accepting an MC move,
as follows.
If the change in the unperturbed energy for an MC move is $\Delta E^0$ and the change in the perturbation
is $\Delta V=-\sum_p h_p\Delta n_p$ then for $\Delta E^0\neq0$ we take 
take $P_\mathrm{a}^h = \mathrm{min}(1,\ee^{(\lambda-1)\Delta E^0/T-\Delta V/T})$ while
for $\Delta E^0=0$ we take $P_\mathrm{a}^h = 2\alpha/(1+\ee^{\Delta V/T})$.  It is easily verified
that this choice is compatible with detailed balance and reduces to the unperturbed probabilities
$P_\mathrm{a}$ if $\Delta V=0$.  Further, coupling the field $h_p$ only to the acceptance
probability ensures that the perturbation affects the rates for bond-breaking and bond-making but does
not affect the rates for diffusion of whole clusters.  

We use a straightforward generalisation of 
the `no-field' method~\cite{Chatelain2004,Ricci-Tersenghi2003} to allow efficient measurement of
the response: see Appendix~\ref{app:chat} for details.
To attain good statistics, we consider responses in which the perturbing field $h$ acts
not just for one MC move but for a time interval $\Delta t=10$ MC sweeps.  Since we are working at leading order in $h$, 
the response to such a perturbation is simply
\begin{equation}
R_{\Delta t}(t,w) \equiv \frac{\delta t}{\Delta t} \sum_{j=0}^{(\Delta t/\delta t)-1} R(t,w+j\delta t).
\end{equation}

The relevant correlation function in this case is $C(t,w)=\langle \delta n_p(t) \delta n_p(w) \rangle$ where
$\delta n_p(t)=n_p(t)-\langle n_p(t)\rangle$ as usual.  It is convenient to define normalised correlation
and response functions
\begin{align}
\tilde{S}_{\Delta t}(t,w) &= \frac{1}{{\cal N}_{\Delta t}(w)} [C(t,w+\Delta t)-C(t,w)]
\label{equ:stilde}
\\
\tilde{R}_{\Delta t}(t,w) &= \frac{1}{{\cal N}_{\Delta t}(w)} R_{\Delta t}(t,w)
\end{align}
with ${\cal N}_{\Delta t}(w) = C(w+\Delta t,w+\Delta t)-C(w+\Delta t,w)$ so that $\tilde{S}_{\Delta t}(w+\Delta t,w)=1$.  We 
also define
\begin{equation}
\tilde{\Delta}_{\Delta t}(t,w) = \tilde{S}_{\Delta t}(t,w)- \tilde{R}_{\Delta t}(t,w).
\end{equation}
In all cases, the subscript $\Delta t$ indicates that these functions generalise the $R(t,w)$, $S(t,w)$ and $\Delta(t,w)$ of
the previous section, while the tilde indicates normalisation. 
For small $\Delta t$, the dependence of these functions on $\Delta t$ is weak, but numerical calculation of the response is easier
for larger $\Delta t$.


%
In Fig.~\ref{figsr} we show numerical results for correlation and response functions.
At $\Teb=0.5$,  clusters of particles are growing in the system, which is far from global equilibrium.  However,
as found in~\cite{Jack2007,Klotsa2011} for several other self-assembling systems, the deviations from FDT are
small because the
time-evolution is nearly reversible (recall Fig.~\ref{figft}). 
On the other hand, at $\Teb=0.15$, the particles are aggregating in disordered clusters and
the time evolution is far from reversible with unbonding events being rare.  

The potential utility of this result was discussed in~\cite{Jack2007,Klotsa2011}: it means that straightforward
measurements on short time scales can be used to predict long-time assembly yield, by exploiting
links between correlation-response measurements and reversibility.  In what follows, we explore
in more detail how the deviation function $\Delta$ couples to microscopic irreversibility during 
assembly.

\subsection{Interpretation of $\Delta(w+\delta t,w)$ as a flux}
\label{sec:zflux}

\begin{figure}
\includegraphics[width=7cm]{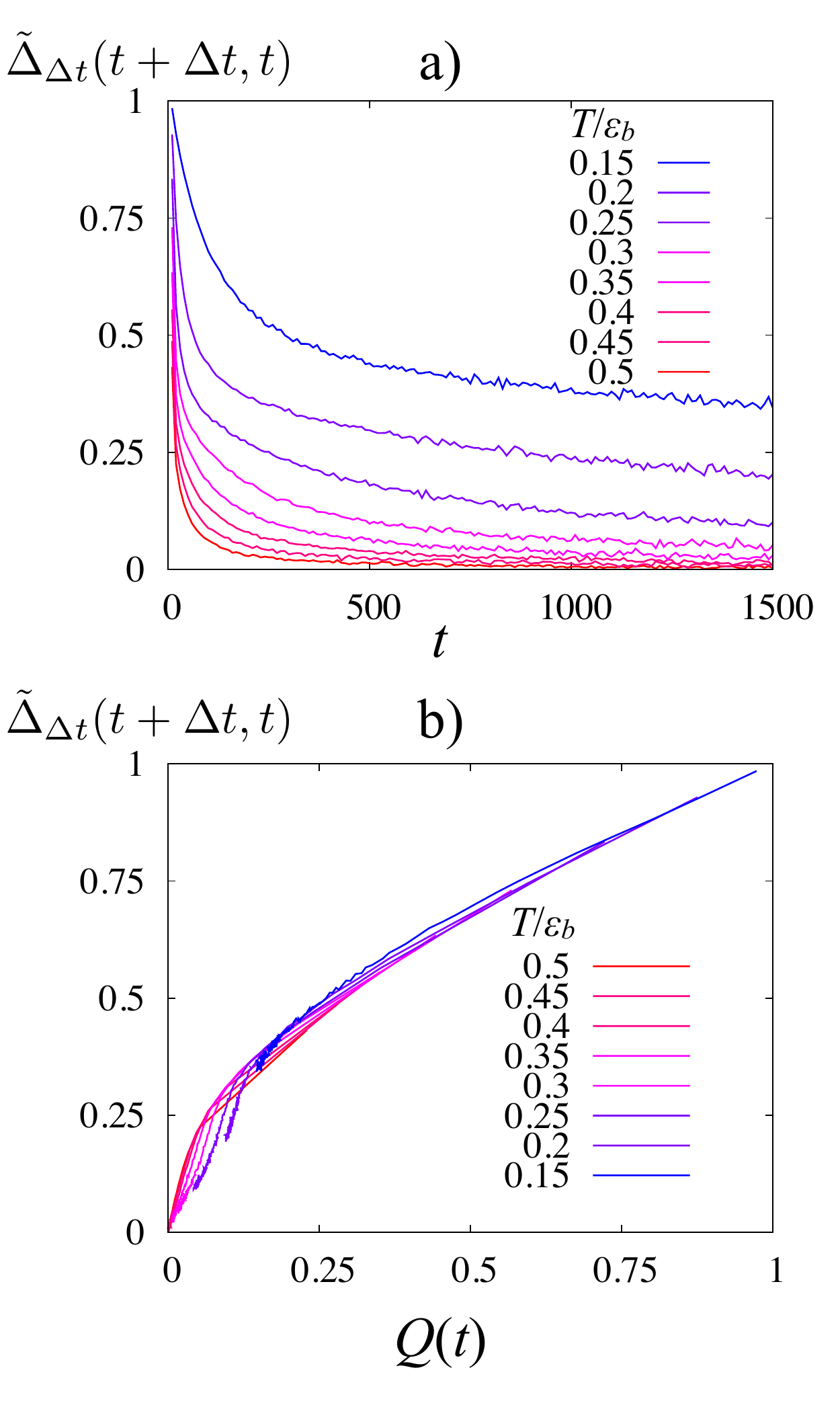}
\caption{\change{
a) The deviation from FDT measured immediately after perturbation, $\tilde{\Delta}_{\Delta t}(t+\Delta t,t)$, $\Delta t=10$.  
At temperatures above $T^*$, $\tilde{\Delta}_{\Delta t}(t+\Delta t,t)$ decreases quickly with $t$, while at low temperatures it remains significant throughout the early times.  In the construction of $\Delta(t,w)$ we take the observables to be $A=B=n_p$ as discussed in the main text.
(b) Parametric plot of $\tilde{\Delta}_{\Delta t}(t+\Delta t,t)$ against the flux-traffic ratio $Q(t)$, for $0<t<1500$.  
The relationship between these two quantities depends very weakly on temperature, indicating the close relation between them.}}
\label{figkvsr}
\end{figure}

In this section, we concentrate on quantities that can be written in the form
\begin{equation}
Z(t) = \sum_{\mu\nu} z(\nu,\mu) J^0_t(\nu,\mu)
\label{equ:ZZ}
\end{equation}
where $J^0_w$ is the probability current, defined in (\ref{equ:j0}).
In contrast to straightforward one-time observables like $\langle A(t) \rangle = \sum_\mu A(\mu) \rho_t(\mu)$, we will
show that observables like $Z(t)$ are currents, time-derivatives and fluxes: they 
measure deviations from instantaneously reversible behaviour, measured at time $t$.

To this end, we generalise the analysis of ``kinks'' in Sec.~\ref{sec:flux-traffic} 
by defining $\langle K^{\nu\mu}(t,\Delta t) \rangle$ as the average 
number of MC transitions from state $\mu$ to state $\nu$ between times $t$ and $t+\Delta t$.  The associated kink rate 
is $k^{\nu\mu}(t) = (\delta t)^{-1} \langle K^{\nu\mu}(t,\delta t) \rangle$.  It follows from the definition of the $k$s that
\begin{equation}
k^{\nu\mu}(t) - k^{\mu\nu}(t) = J^0_t(\nu,\mu)
\end{equation}
%
%
Thus, the probability current $J^0_t$ gives the difference between the probabilities of forward and reverse MC transitions between states $\mu$ and $\nu$,
which clarifies its connection to irreversibility.  At equilibrium,
one has $J^0_t(\nu,\mu)=0$ for all states $\mu$ and $\nu$: thus, $J^0_t$ measures deviations from equilibrium. 
However, measuring (or even representing) $J^0_t$ is a near-impossible task in a system where the number of possible states $\mu$ is
exponentially large in the system size. 

Instead, one makes a specific choice of the 
$z(\nu,\mu)$, in which case (\ref{equ:ZZ}) shows that $Z(t)$ is a projection of the current $J^0_t$ onto the observables of the system.
For example, if $z(\nu,\mu) = A(\nu)$ then $Z(t) = \frac{\partial}{\partial t}\langle A(t) \rangle$ is a time derivative that
clearly vanishes at equilibrium (or in any steady state).
Other choices for $z(\nu,\mu)$ become relevant in out-of-equilibrium settings.  For example, the flux $f(t)$ in Sec.~\ref{sec:flux-traffic}
is obtained by setting $z(\nu,\mu)=1$ if the transition $\mu\to\nu$ involves an increase in the number of bonds of particle $p$,
with $z(\nu,\mu)=0$ for all other $\mu,\nu$. (The structure of (\ref{equ:ZZ}) together with this choice of $z$ ensures that 
$f(t)$ acquires negative contributions from MC transitions where particle $p$ experiences a decrease in its number of bonds, as required.)
Similarly the particle current in exclusion processes
is obtained by taking $z(\nu,\mu)=1$ for transitions $\mu\to\nu$ where a particle hops to the right, and zero for all
other pairs.  Clearly, these observables monitor the breaking of time-reversal symmetry, as is the case for all observables 
$Z(t)$. 

To relate deviations from FDT to irreversible events, we first consider the case $t=w+\delta t$, so that the
perturbation $h$ is applied for just one MC move and then the response is measured immediately.
In this case (\ref{eqdeviation}) reduces to  
\begin{equation}
\Delta(w+\delta t,w) = \sum_{\nu,\mu} z_\Delta(\nu,\mu) J_w^0(\nu,\mu)
\label{equ:delta-immed}
\end{equation}
as in (\ref{equ:ZZ}), with
\begin{equation}
z_\Delta(\nu,\mu) = [ A(\nu) - \langle A(w+\delta t) \rangle  ] F(\nu,\mu).
\label{equ:z_immed}
\end{equation}
The function $F(\nu,\mu)$
was defined in (\ref{equ:Fnm-general}) and measures the effect of the perturbation $h$ on the transition rate from
$\nu$ to $\mu$.  The factor $A(\nu) - \langle A(w+\delta t) \rangle$ indicates whether configuration $\nu$ has a high
or a low value of the observable $A$, compared to the average of $A$ at the measurement time $w+\delta t$.

For the case considered in Sec.~\ref{sec:resp-energy} and Fig.~\ref{figsr}, where the perturbation is coupled
to the energy of particle $p$, one has $A(\nu) = n_p(\nu)$ and
\begin{equation}
F(\nu,\mu) =[ n_p(\nu)-n_p(\mu) ] \Theta(E^0(\mu)-E^0(\nu))
\label{equ:Fnm}
\end{equation}
where $\Theta(x)$ is the step function, defined such that $\Theta(0)=\frac12$.
The step function appears because if a proposed MC move results in a decrease in the bare energy, 
the acceptance probability does not depend on the perturbing field $h_p$.  Clearly,
$F(\nu,\mu)$
is finite only if particle $p$ changes its energy between states $\mu$ and $\nu$: the same is true
for $z_\Delta(\mu,\nu)$. 

 Thus, (\ref{equ:delta-immed}) indicates that the
deviation $\Delta(t,w)$ measured at $t=w+\delta t$ reflects the imbalance 
between rates of bond-making and bond-breaking for particle $p$.  In this respect it is similar to the flux $f(w)$: however,
the weights given to different bond-making and bond-breaking processes differ between $\Delta(w+\delta t,w)$ and $f(w)$, due to
the different forms of $z(\nu,\mu)$ in the two cases.  Thus, while these quantities reflect similar physics, the details of their
behaviour is different.
%
To see the relationship between the immediate response $\Delta(w,w+\delta t)$ and the 
flux $f(t)$ from Sec.~\ref{sec:flux-traffic}, we present Fig.~\ref{figkvsr} which shows
that the two quantities have similar time and temperature-dependence and therefore reveal
similar information about the (ir)reversibility of the dynamical bonding and unbonding processes.


\subsection{Time intervals for reversibility}

\begin{figure}
\includegraphics[width=7cm]{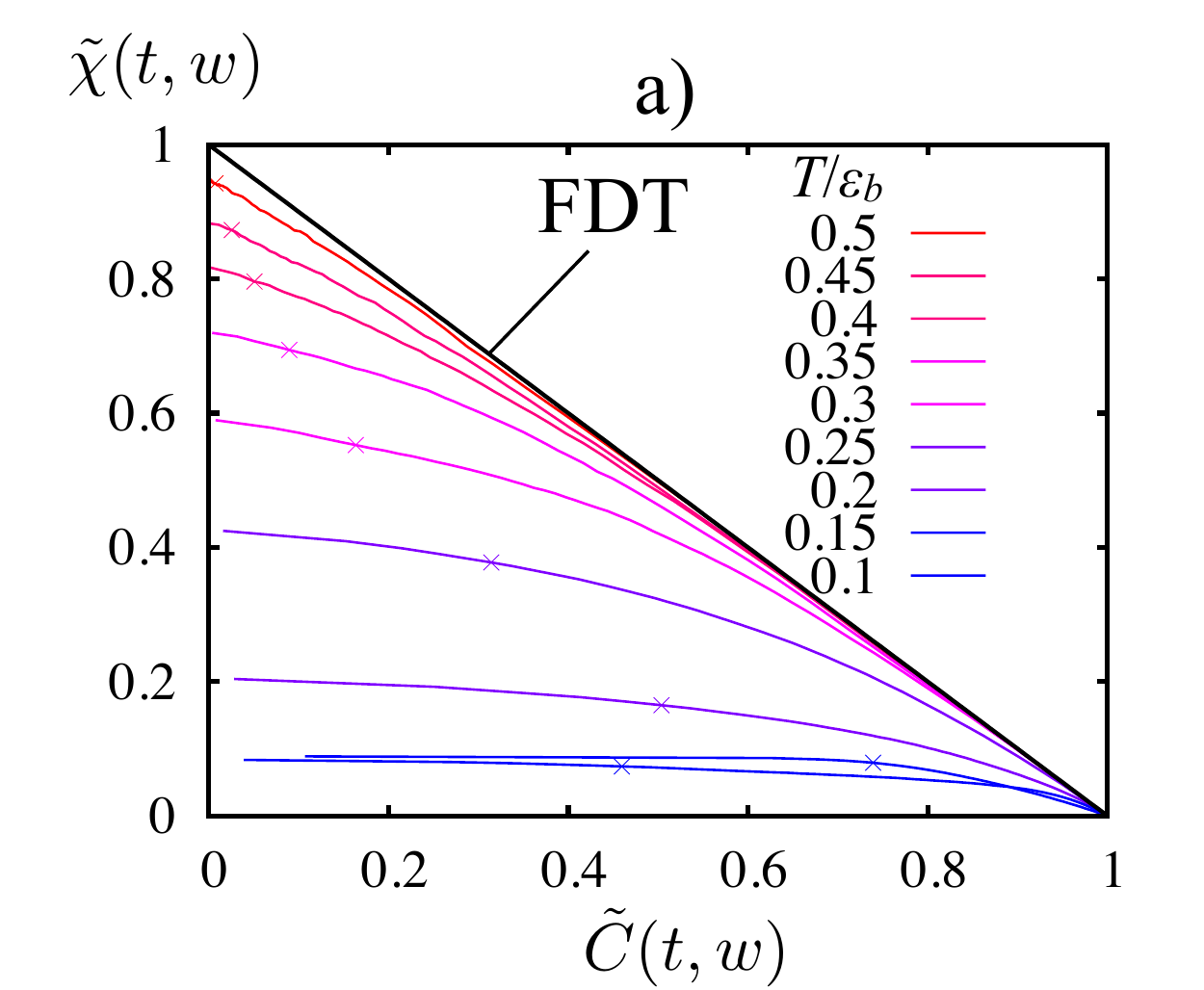}
\caption{
Integrated correlation-response plot showing the response $\tilde\chi(t,w)$ and correlation $\tilde{C}(t,w)$.  The correlation
and response functions are associated with observables $A=B=n_p$, as in previous figures.  The data is shown for a fixed
 $t=10^4$ MCS and and $0\leq w < t$.  The $\times$ symbol indicates the points where $w=10^3$: impulse responses associated with this time
 are shown in Fig.~\ref{figsr}.}%
\label{figfdt}
\end{figure}

We now consider the deviation $\Delta(t,w)$ for $t>w+\delta t$.  
From (\ref{eqdeviation}), we see that $\Delta(t,w)$ 
can be written in the form (\ref{equ:ZZ}) if we take
\begin{equation}
z(\nu,\mu) = [ \mathcal{P}^A_u(\nu) - \langle A(t) \rangle  ] F(\nu,\mu).
\label{equ:z_prop}
\end{equation}
where
\begin{equation}
{\cal P}^A_{u}(\nu)=\sum_\gamma 
A(\gamma) G^0_u(\gamma\leftarrow\nu)
\end{equation}
is the propensity~\cite{Harrowell2004} of observable $A$ for configuration $\nu$ after a time $u$.  That is,
${\cal P}^A_{u}(\nu)$ is the value of $A$ that is obtained by an average over the dynamics of the system,
for a fixed initial condition $\nu$ and a fixed time $u$.  (Recall $u=t-w-\delta t$.)  

Comparing (\ref{equ:z_prop}) with (\ref{equ:z_immed}), the 
difference is the replacement of $A(\nu)$ by the propensity.  At equilibrium, one expects
$\mathcal{P}^A_u(\nu) - \langle A(t) \rangle$ to decay towards zero as $u$ increases, since the system forgets its memory
of the configuration $\nu$ and ``regresses back to the mean''. (That is, the propensities for different fixed initial
conditions all converge to the same average value at long times.)  
Out of equilibrium, this may not be the case: if a configuration
$\nu$ has a more ordered structure than another configuration $\nu'$ then these configurations may have different
 propensities for order even at much later times.  Loosely, a system in $\nu$ has a `head start' along the route to assembly,
 and the system does not forget the memory of this head start as assembly proceeds.

Fig.~\ref{figsr} shows that for the correlation and response functions considered here, the deviation $\Delta(t,w)$ decays only
very slowly with time $t$.  The $t$-dependence of this function comes through a weighted sum of propensities: it is therefore
clear that these propensities do not regress quickly to the mean.  In contrast, one may write the correlation as
\begin{multline}
S(t,w) = \sum_{\nu\mu} [ {\cal P}_u^A(\nu) - \langle A(t) \rangle ]     [ n_p(\nu) - n_p(\mu) ] \\ \times W^0(\nu\from\mu) \rho_w(\mu)
\end{multline}
whose $t$-dependence also comes from the propensity.  Fig.~\ref{figsr} shows that this function does decay quite
quickly with $t$, although it does not reach zero on the time scales considered here.

Our conclusion is the following.  For configurations $\nu$ that are typical at time $w$, 
the propensity ${\cal P}_u^A(\nu)$ has a `fast' contribution that decays quickly
with time $u$, as well as a `slow' contribution that depends weakly on $u$ and
reflects the `head start' of $\nu$ along the route to the assembled product.  The fast contribution reflects
rapid bond-making and bond-breaking events that do not lead directly to assembly while the `slow' contribution
is a non-equilibrium effect that measures assembly progress and also dominates the deviation $\Delta(t,w)$ 
that we have defined here.  By contrast $S(t,w)$ picks up contributions from both slow and fast contributions.  The utility of the FDR is that for suitable $w$, it couples to the irreversible bond-making
behaviour that is most relevant for self-assembly.

To illustrate this balance of fast and slow degrees of freedom, and also 
to make contact with previous studies~\cite{Crisanti2003,Jack2007,Klotsa2011}
of FDRs, we define an integrated (and normalised)
response
\begin{equation}
\tilde\chi(t,w)=\frac{1}{C(t,t)}\int_w^t \!\mathrm{d}w' R(t,w')
\end{equation}
and we also normalise the correlation as
\begin{equation}
\tilde C(t,w)=\frac{C(t,w)}{C(t,t)}
\end{equation}

It is conventional to display these correlations functions in a parametric form~\cite{Crisanti2003}.  
Specifically, making a parametric plot of $\tilde\chi(t,w)$ against $\tilde C(t,w)$, for fixed $t$ and varying $w$,
the gradient is $-X(t,w)$.  
\change{It has been emphasised several times~\cite{Sollich2002,Jack2006,Russo2010} that plotting
data for fixed $w$ and varying $t$ is not in general equivalent to this procedure and may give misleading results,
especially if $C(t,t)$ has significant time-dependence, as it does in these systems.}  
Plotting data at fixed $t$ as in Fig.~\ref{figfdt}, it is apparent that the high temperature systems are the most
reversible, with $X(t,w)\approx 1$.  To understand the connection with Fig.~\ref{figsr}, note that
the time interval $t-w$ increases from right to left in Fig.~\ref{figfdt}: 
when $t-w$ is large then the deviation $\Delta(t,w)$ 
has a larger fractional contribution to $S(t,w)$, so that the curves are steepest (most reversible)
when $t-w$ is small and least steep (least reversible) when $t-w$ is large.
\change{We find that parametric plots depend weakly on the fixed values of $t$ used but we do not
show this data, for brevity:
see for example Ref.~\cite{Klotsa2011} for an analysis of $t$-dependence in a system of crystallising particles.}

Physically, we attribute the $w$-dependence of $X(t,w)$ to the fact that fast degrees of freedom (like 
dimer formation and breakage in the vapour phase) quickly relax to a quasiequilibrium
state~\cite{Hagan2011} where probability currents $J^0$ associated with this motion are small.  It
is only much slower degrees of freedom (like the gradual growth/assembly of large clusters)
for which probability currents remain significant on long time scales, and lead to
long-time contributions to $\Delta(t,w)$.

Compared to the simple flux-traffic ratio $Q(t)$ considered in Sec.~\ref{sec:flux-traffic}, the normalised
deviation $1-X(t,w) = \Delta(t,w)/S(t,w)$ has the same effect of comparing a generalised flux $\Delta(t,w)$ with
a normalisation $S(t,w)$ that reflects the traffic in the system.  However, the effect of the time difference $t-w$
is that while $\Delta(t,w)$ decays slowly with time $t$ then $S(t,w)$ decays quite quickly.  Thus, as $t$ increases, the deviation
$1-X(t,w)$ becomes increasingly sensitive to the deviations from reversibility, since the fluctuations of fast (reversible) degrees of freedom
regress back to the mean while the slow (irreversible) ones remain significant.  For this reason, the FDR $X(t,w)$ is a more
sensitive measure of irreversibility than the flux-traffic ratio discussed in Sec.~\ref{sec:flux-traffic}.

\subsection{Responses to different perturbations}


\label{sec:site}
\begin{figure}
\includegraphics[width=7cm]{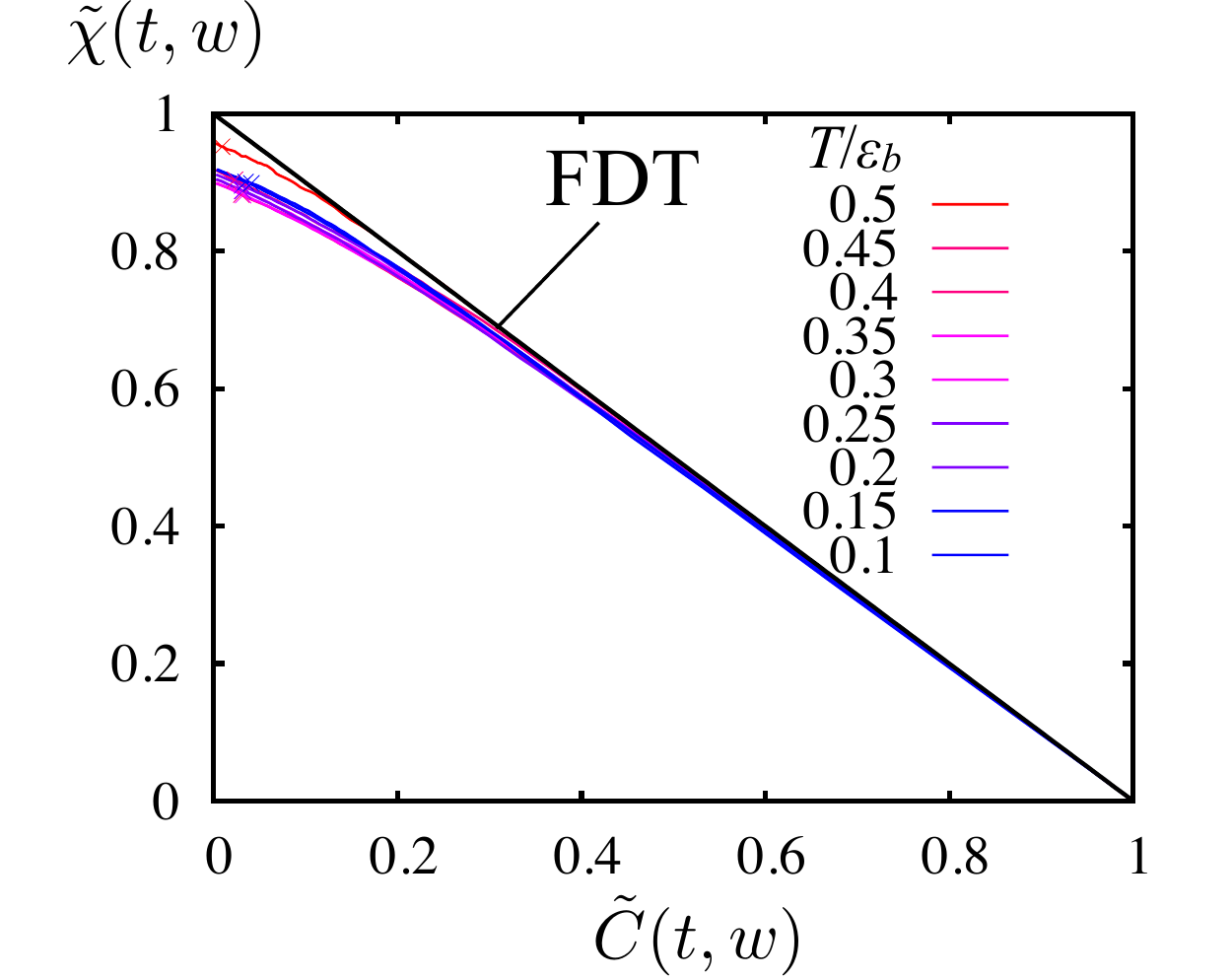}
\caption{
Integrated correlation-response plot a site dependent perturbation. That is, we take the observables $A=B=\rho_i$ as described Sec.~\ref{sec:site},
and in contrast to previous figures where $A=B=n_p$.
The data is shown for fixed $t=10^4$ with $\times$~symbols indicating $w=10^3$, as in Fig.~\ref{figfdt}.  
The relaxation of the observable $\rho_i$ is diffusive and depends weakly on the bond strength, so
the dependence on $\Teb$ is much weaker than in Fig.~\ref{figfdt} and the deviations from FDT behaviour are smaller.
}
\label{figfdtobs}
\end{figure}

We remark that the observables used in measuring correlation and
response functions may strongly affect the results.  Most relevant is
the extent to which the response function couples to the slow, irreversible
degrees of freedom and the fast, reversible ones.  To illustrate this,
we compare the results of the previous section with a different response function.
We take the observables $A$ and $B$ of Sec.~\ref{sec:flux-relation} both equal
to the occupancy $\rho_i=0,1$ of a specific site $i$ on the lattice.  The effect
of the perturbation on the MC dynamics is the same as that described in Sec.~\ref{sec:resp-energy},
except that the perturbed energy $E^h=E^0+V$ with $V=\sum_i h_i \rho_i$, where the sum runs over
sites of the lattice.

Unlike the bond numbers $n_p$ which change only when bonds
are made and broken, the site occupancies $\rho_i$ also change as
clusters diffuse through the system.  When bonds are strong, the site
correlation functions decay faster than the bond correlation functions
considered above.  Therefore, since the site response is coupling to fast degrees
of freedom, we expect to see `more reversible' behaviour: this is borne out by 
Fig.~\ref{figfdtobs} where the FDRs are much closer to the equilibrium value (unity) 
than the FDRs shown in Fig.~\ref{figfdt}.  This is consistent with recent
results of Russo and Sciortino~\cite{Russo2010} who observed a similar effect, and with
earlier studies of observable-dependence of FDRs~\cite{Sollich2002,Mayer2003,Jack2006} 
(see however \cite{Mayer2005} in which some fast degrees of freedom do not relax to quasiequilibrium).

Similar correlation and response measurements in an Ising lattice gas were also considered by 
Krz\k{a}ka\l{}a \cite{Krzakala2005}, but the dynamical scheme used in that work means that clusters of 
particles do not diffuse: in that case the site degrees of freedom $\rho_i$ are not able to reach quasi-equilibrium 
and one does observe significant deviations from FDT for this observable.

\change{
We end this section by briefly considering these correlation and response functions for large $t$.
In this limit, standard arguments~\cite{Crisanti2003} indicate that both correlation and response functions have contributions from well-separated fast
and slow sectors, with the contributions from the fast sector obeying FDT.  In coarsening systems like this one then there is no response
in the slow sector.  Thus, for large enough times, the parametric plot has a segment with $X=1$ for large $C$ and small $\chi$; for
smaller $C$ then $X=0$ so the parametric plot has a plateau where the value of $\chi(t,w)$
depends only on equilibrium properties of the system.  However, we emphasise that the approach to this large-$t$ limit is very slow and applies
only after all clusters in the system are annealed into compact clusters.  This limit is therefore irrelevant for analysing the kinetic
trapping phenomena on which we focus in this article.
}

\subsection{Relation to previous analyses of non-equilibrium response}
\label{sec:lippi-etc}
%


We have emphasised throughout this article that our main purpose is to use correlation and response measurements to
measure reversibility in assembling systems.  Many other studies have considered such measurements in a variety of
other contexts -- here we make connections between our methodology and some other results from the literature.

One recent area of
interest has been the use of `no-field' measurements of response functions.  Here, the aim is to develop formulae for the response
that can be evaluated in Monte Carlo simulations without the introduction of any direct perturbation~\cite{Chatelain2004,
Ricci-Tersenghi2003,Lippiello2005,Jack2006,Corberi2010}.
In fact, we use such a straightforward generalisation of the method of~\cite{Chatelain2004} to measure responses in this paper,
as discussed in Appendix~\ref{app:chat}.

Our analysis in Sec.~\ref{sec:flux-relation} is concerned not so much with measurement of the response, but with
the physical interpretation of deviations from FDT.  Our result is therefore in the same spirit as~\cite{Cugliandolo1997,Cugliandolo1997a,
Crisanti2003,Lippiello2005,Jack2007,Baiesi2009,Seifert2010}.  In particular, Baiesi~\emph{et al}~\cite{Baiesi2009} 
relate the FDT to `flux' and `traffic' observables that separate reversible and irreversible behaviour, 
but they use a different approach to ours.  Full details are given in Appendix.~\ref{app:maes} but in our notation, their central result is
\begin{equation}
R(t,w) = \frac12 \left[  \frac{\partial}{\partial w} \langle A(t) B(w) \rangle + \langle A(t) \mathcal{T}(w) \rangle \right]
\label{equ:maesfinal}
\end{equation}
where $\mathcal{T}(w)$ measures the $h$-dependence of the amount of dynamical activity (traffic) between times $w$ and $w+\delta t$.
The second term is therefore associated with traffic while the first term is a correlation between $A(t)$ 
and the `excess entropy production' at time $w$: it is therefore related to a flux. 
However, neither of the correlation functions in (\ref{equ:maesfinal}) may
 be written in the form of (\ref{equ:ZZ}) so they do not vanish at equilibrium.
The key point is
that in equilibrated (reversible) systems, the terms in (\ref{equ:maesfinal}) are both non-zero and equal to each other.  So while
(\ref{equ:maesfinal}) does involve a separation of terms symmetric and anti-symmetric under time reversal, the resulting correlation
functions are not fluxes in the form of (\ref{equ:ZZ}) and do not provide the same direct measurement of the deviation from reversibility
given by $\Delta(t,w)$ in (\ref{eqdeviation}).

In another relation to irreversible behaviour, 
the analysis of Refs.~\cite{Lippiello2005,Corberi2010} identifies an `asymmetry' measurement which in our notation is
\begin{equation}
\mathcal{A}(t,w) = R(t,w) - \frac12[ S(t,w) + S(w,t) ]
\end{equation}
It follows from the fluctuation dissipation theory that this quantity vanishes in a system with time-reversal symmetry. 
Comparison with (\ref{equ:def-Delta}) also illustrates the similarity between $\mathcal{A}(t,w)$ and the deviation $\Delta(t,w)$.
However,
$\mathcal{A}(t,w)$ differs from $\Delta(t,w)$ since the deviation $\Delta(t,w)$ is a projection of $J^0_w$ and hence vanishes if the system is behaving
reversibly at time $w$, regardless of what happens at later times.  On the other hand,
$\mathcal{A}(t,w)$ vanishes only if all trajectories between times $w$ and $t$ are equiprobable with 
their time-reversed counterparts. That is, time-reversal symmetry must hold throughout the interval between $w$ and $t$, and not just
at time $w$.

\section{Outlook}

We have analysed the extent to which ideas of reversible bond-making are applicable
to a phase separating lattice gas, which we used as a simplified model for a self-assembling
system.  The data of Figs.~\ref{figyld}, \ref{figft} and~\ref{figfdt} further support the idea
that this system is relevant for studying self-assembly, since similar results have been
presented for more detailed models of self-assembling 
systems~\cite{Hagan2006,Jack2007,Rapaport2008,Klotsa2011,Grant2011}.

The flux/traffic measurements of Sec.~\ref{sec:flux-traffic} provide an intuitive picture 
of reversibility, while the picture based on the correlation-response formalism is more
subtle.  However, the central result (\ref{eqdeviation}) demonstrates an
explicit link between measurements of response functions and irreversibility of bonding.
Also, Figs.~\ref{figsr}
and~\ref{figfdt} do indicate that information about both short-time reversible and
long-time irreversible behaviour can be obtained by considering the behaviour of
correlation and response functions.

It has been argued that correlation-response measurements on short time scales
might be used to predict long-term assembly yield, and even to control assembly
processes.  The work presented here places this objective on a firmer theoretical
footing, and highlights the importance of using the right observable when probing
irreversibility (compare Figs.~\ref{figfdt} and~\ref{figfdtobs}); the interplay
between fast `quasiequilibrated' degrees of freedom and slow `irreversible' bonding~\cite{Hagan2011}
also highlights the importance of choosing the right time scale for measuring
these functions and predicting long-time assembly quality.

\begin{acknowledgments}
We thank Mike Hagan, David Chandler, Steve Whitelam and Paddy Royall for many useful discussions on
reversibility in self-assembly.  We gratefully acknowledge financial support by the EPSRC through a doctoral training grant 
(to JG) and through grants EP/G038074/1 and EP/I003797/1 (to RLJ).
\end{acknowledgments}

\begin{appendix}

\section{Derivations of representations for response functions}
\label{app:response}

\subsection{Deviation function $\Delta(t,w)$}
\label{app:dev}

In this Section, we derive equation (\ref{eqdeviation}) which gives the deviation between 
the response $R(t,w)$ and the correlation function $S(t,w)$.  Starting from (\ref{eqr1}),
conservation of probability for the MC transition probabilities implies that $P^h(\mu\from\mu) = 1 - \sum_{\nu(\neq\mu)} P^h(\nu\from\mu)$,
with a similar relation for the rates $W^h$.  Using this in (\ref{eqr1}) gives 
\begin{equation}
R(t,w)=T\sum_{\gamma,\nu\neq\mu}A(\gamma)G_u(\gamma\leftarrow\nu) 
\frac{\partial}{\partial h}J^h_w(\mu,\nu)
\label{eqr3}
\end{equation}
with 
\begin{equation}
J^h_w(\mu,\nu) = W^{h}(\nu\leftarrow\mu)\rho_w(\mu)-
W^{h}(\mu\leftarrow\nu)\rho_w(\nu)
\end{equation}
which has the structure of a flux in probability from configuration $\mu$ to configuration $\nu$.
This is related~\cite{Seifert2010} to an expression for the response originally due to Agarwal~\cite{Agarwal1972}.

We now assume that the system obeys detailed balance:
\begin{equation}
W^h(\mu\from\nu) \ee^{-E^h(\nu)/T} = W^h(\nu\from\mu) \ee^{-E^h(\mu)/T}
\label{equ:detbalW}
\end{equation}
where $E^h(\mu) = E^0(\mu) - hB(\mu)$ as defined in Sec.~\ref{sec:flux-relation}.
Hence, one has $\frac{\partial}{\partial h} \ln W^h(\nu\from\mu) = \Delta B(\nu,\mu)/T +
 \frac{\partial}{\partial h} \ln W^h(\mu\from\nu)$ with $\Delta B(\nu,\mu)=B(\nu)-B(\mu)$, so that 
\begin{multline}
T\frac{\partial}{\partial h}J^h_w(\mu,\nu) = 
\Delta B(\nu,\mu) W^0(\nu\from\mu) \rho_w(\mu)
\\ + 
TJ^0_w(\nu,\mu) \frac{\partial}{\partial h} \ln W^h(\mu\from\nu).
\label{equ:JhJ0}
\end{multline}

Inserting (\ref{equ:JhJ0}) into (\ref{eqr3}) gives two contributions to the response.  The first
contribution is $\frac{1}{\delta t}\langle A(t) [ B(w+\delta t)-B(w) ]\rangle$.  In the notation
of the main text, this is equal to 
$S(t,w)+\langle A(t)\rangle \frac{\partial}{\partial w}\langle B(w)\rangle$ where the time derivative
is interpreted as a change between $w$ and $w+\delta t$, normalised by the time $\delta t$ for a single MC move. (Recall that we
took time
to be discrete in these MC models.)  

At equilibrium, $J^0=0$ and $\frac{\partial}{\partial w}\langle B(w)\rangle=0$ so the
only contribution to the response is $S(t,w)$ and the FDT (\ref{equ:fdt}) holds.  The non-equilibrium contributions to $R(t,w)$ can be collected together to obtain (\ref{eqdeviation})
if one notes (i) that that $\sum_\gamma G^0_u(\gamma\from\nu)=1$ which follows from conservation
of probability and the definition of $G^0$ and (ii) that
\begin{equation}
\sum_{\mu\nu} F(\nu,\mu) J^0_w(\nu,\mu)=\frac{\partial}{\partial w}\langle B(w)\rangle,
\end{equation} which follows
from the detailed balance property of the $W^h$.

\subsection{`No-field' formula for the response}
\label{app:chat}

Another useful representation of the response function expresses it in terms of directly observable 
quantities in unperturbed simulations~\cite{Chatelain2004,Ricci-Tersenghi2003} (this is the `path weight'
representation of~\cite{Seifert2010}).  

The original `no-field' method requires an extension for our system since the cluster
MC algorithm we use means that the same transition $\nu\from\mu$ may take place via several possible
computational routes. (For example, two MC moves may start from different seed particles but result
in the same cluster being moved in the same direction).  We use $P(\nu\xleftarrow{C}\mu)$ to represent
the probability of an MC move from $\mu$ to $\nu$ by some route $C$.  For $\nu\neq\mu$, the route
$C$ is the choice of seed particle and sequence of steps by which the moving cluster is generated.
If $\nu=\mu$ the route $C$ 
may involve the proposal of a move that is then rejected, or the proposal of a move where the cluster size exceeds $n_\mathrm{max}$, or would result in multiple-occupancy of a site.  

Starting from (\ref{eqr1}), one writes 
\begin{multline}
R(t,w)=T\sum_{\gamma,\nu,\mu,C}A(\gamma)G^0_u(\gamma\from\nu) \mathcal{R}(\nu,C,
\mu) \\ \times P^0(\nu\xleftarrow{C}\mu)\rho_w(\mu).
\end{multline}
where $\mathcal{R}(\nu,C,\mu) = \frac{1}{\delta t} \frac{\partial}{\partial h}\log W^h(\nu\xleftarrow{C}\mu)$.
This result may be written as a simple expectation value
\begin{equation}
R(t,w)=T\langle A(t) \mathcal{R}(w) \rangle
\label{equ:chat}
\end{equation}
where the key point is that the observable ${\cal R}(\nu,C,\mu)$ may be evaluated for any attempted
MC move so that the correlation function in (\ref{equ:chat}) may be calculated directly
from an MC simulation with $h=0$.  
[This is in contrast to the current $J^0_w(\nu,\mu)$ which depends on the whole ensemble of
evolving systems, so that the formula (\ref{eqr3}) cannot be evaluated directly by MC simulation.]


\comment{For the dynamical method described in this paper, the observable ${\cal R}(\nu,C,\mu)$ is given by ${\cal R} =[ n_p(\nu)-n_p(\mu) ] \Theta(E^0(\mu)-E^0(\nu))$ if the move is accepted, as in the main text $\Theta(0)=\frac{1}{2}$.  If the move is rejected either because the cluster size $n$ exceeds $n_{\rm max}$ or the move would have resulted in overlapping particles  ${\cal R} =0$.  If however the move is legal but rejected when testing the move acceptance probability $P_{\rm a}$, then ${\cal R} =-[n_p(\nu)-n_p(\mu) ] \Theta(E^0(\mu)-E^0(\nu))\times P_{\rm a}/(1-P_{\rm a})$.}

\subsection{An alternative formula for the response}
\label{app:maes}


In this section we clarify the connections between our results and those of Baiesi \emph{et al.}~\cite{Baiesi2009}.
The central identity in the derivation of Appendix~\ref{app:dev} is $\frac{\partial}{\partial h} \ln W(\nu\from\mu) =
 \Delta B(\nu,\mu)/T + \frac{\partial}{\partial h} \ln W(\mu\from\nu)$.  Rearranging, one may write
 \begin{multline}
 \frac{\partial}{\partial h} \ln W(\nu\from\mu) =
 \Delta B(\nu,\mu)/2T  \\ + \frac12 \frac{\partial}{\partial h} \ln [ W(\mu\from\nu) W(\nu\from\mu)]
 \end{multline}
and substitution into (\ref{eqr1}) leads to
\begin{multline}
 R(t,w) = \frac12 \sum_{\gamma\mu\nu} A(\gamma) G_u(\gamma\from\nu) \\ \times
[  \Delta B(\nu,\mu) + {\cal T}(\nu,\mu) ]  W(\nu\from\mu) \rho_w(\mu)
\label{equ:maes}
\end{multline}
with ${\cal T}(\nu,\mu) = T\frac{\partial}{\partial h} \ln [ W(\mu\from\nu) W(\nu\from\mu)]$.  Equ.~(\ref{equ:maes})
is the translation of the central result of Baiesi~\emph{et al}.~\cite{Baiesi2009} into our notation.

The key point here is that the two terms in square brackets in (\ref{equ:maes}) have different symmetry properties.
By definition of the first term, $\Delta B(\nu,\mu)=-\Delta B(\mu,\nu)$.  Time-reversal of the trajectory
flips the roles of $\nu$ and $\mu$ so that $\Delta B$ is `anti-symmetric under time-reversal' and is related to
the entropy current~\cite{Baiesi2009}.  This means that
the average of $\Delta B$ is zero in a time-reversal symmetric system. (Note however that $\Delta B$ itself is a fluctuating
quantity and not zero
in general: this is different from the probability current $J^0_w$ which is an ensemble-averaged property without fluctuations 
and is
strictly equal to zero if time-reversal holds.)  On the other
hand ${\cal T}(\nu,\mu)={\cal T}(\mu,\nu)$ so ${\cal T}$ is `symmetric under time-reversal' and measures dynamical
activity, or traffic.

Despite the links between Equ.~(\ref{equ:maes}) and time-reversal symmetries, neither
of the terms in (\ref{equ:maes}) may be written as a flux in the form of (\ref{equ:ZZ}).  Noting that 
$S(t,w)$ is a connected correlation function, one may write
\begin{equation}
 R(t,w) = \frac12 \left[ S(t,w) + \langle A(t) \mathcal{T}(w) \rangle + \langle A(t) \rangle  \frac{\partial}{\partial w} \langle B(w) \rangle \right]
\label{equ:maes2}
\end{equation}
where the final term vanishes in the non-equilibrium steady states considered in~\cite{Baiesi2009} while $\mathcal{T}(w)$ is given
by $(\delta t)^{-1} {\cal T}(\nu,\mu)$, evaluated with $\nu$ and $\mu$ being the configurations at times $w+\delta t$ and $w$ respectively.  The key point
is that the first two terms on the right hand side of (\ref{equ:maes2}) are equal to each other and both are non-zero at equilibrium.
Hence neither of them
is directly identifiable as a flux that is sensitive only to deviations from reversibility.  Combining the first and third terms on the right hand
side of (\ref{equ:maes2}) yields (\ref{equ:maesfinal}) of the main text.

\end{appendix}

\bibliography{biblog-edited}
\end{document}